\documentclass[conference]{IEEEtran}

\usepackage{graphicx}
\usepackage{amsmath}
\usepackage{cite}
\usepackage{booktabs}
\usepackage{xurl}
\usepackage{makecell}
\usepackage{adjustbox}
\usepackage{threeparttable}
\usepackage{ragged2e}
\usepackage{color}
\usepackage{soul}

\begin{document}

\title{SpikeDuet: CPU-Accelerator Co-Execution of Sparse SNNs on Commodity MCUs}

\author{
\IEEEauthorblockN{
Muhammad Fahd Ibrahim Bhatti\IEEEauthorrefmark{1} and Muhammad Ali Siddiqi\IEEEauthorrefmark{1}\IEEEauthorrefmark{4}
}
\IEEEauthorblockA{
\IEEEauthorrefmark{1}Electrical Engineering Department, Lahore University of Management Sciences, Pakistan\\
\IEEEauthorrefmark{4}Department of Quantum and Computer Engineering, Delft University of Technology, The Netherlands\\
}
}
\maketitle

\begin{abstract}
Spiking neural networks (SNNs) expose sparse event activity, but commodity microcontrollers increasingly pair general-purpose processors with dense matrix-vector accelerators whose suitability for sparse inference is unclear. We investigate whether the on-chip Matrix Vector Processor (MVP) of the EFR32MG24 can accelerate SNN inference beyond optimized event-driven Cortex-M33 execution. Our automated snnTorch-to-C deployment flow combines hardware-calibrated engine selection, function-preserving neuron reordering that places co-active neurons adjacently for the accelerator, and asynchronous scheduling that overlaps CPU and MVP computation. On physical hardware, heterogeneous execution achieves 1.27--1.87$\times$ speedup and 11--34\% lower energy per inference than optimized sparse Cortex-M33 execution across five evaluated network configurations, while the smallest Braille network remains CPU-optimal. Reordering alone moves Braille-384's dominant hidden layer from the CPU to the MVP, and its benefit persists under asynchronous execution. These results show that activity locality and concurrency between CPU and MVP can jointly improve sparse SNN inference on commodity MCUs.
\end{abstract}


\section{Introduction}

Spiking neural networks (SNNs) are attractive for edge inference because their computation is driven by sparse spike events: at each timestep, only a subset of neurons fires, so work associated with inactive inputs can be skipped. On microcontrollers (MCUs), this sparsity is exploited by event-driven execution on the central processing unit (CPU). At the same time, commodity MCUs increasingly integrate neural or matrix-vector accelerators alongside the CPU~\cite{millar2025unpu}. These accelerators are designed for dense computation with regular, contiguous memory accesses, which conflicts with the irregular, input-dependent activity of SNNs. Mapped densely, an SNN forfeits the savings of sparsity; mapped sparsely, its computation fragments into many small accelerator invocations whose overheads can outweigh the benefit. It therefore remains unclear whether such an accelerator can speed up SNN inference beyond an optimized sparse CPU implementation.

Among prior approaches to efficient SNN inference, neuromorphic processors execute SNNs natively~\cite{davies2018loihi} but require dedicated silicon absent from commodity MCUs. On general-purpose MCUs, sparse and activity-aware implementations skip computation for inactive inputs~\cite{boretti2023mcu,barchi2021nearsensor,barchi2024activity,liao2024sparsity} but execute only on general-purpose cores. SNN deployment frameworks automate the translation of trained models to embedded C or neuromorphic hardware~\cite{gullett2025embedder,pedersen2024nir,garcia2026snnmlir,jurzec2025compression,pignata2026snn2mcu}, but none targets the dense accelerators found on commodity MCUs. Conversely, toolchains for such accelerators target dense artificial neural networks~\cite{fatima2025purkinje,david2021tflm,lin2020mcunet,liu2024microtvm}, whose computation is regular and input-independent, and therefore cannot exploit spike sparsity. Neuron permutation symmetry and channel reordering have been studied for model merging and pruning~\cite{entezari2022permutation,wan2023upscale}, but not to shape SNN activity for accelerator execution. To our knowledge, no existing toolchain automatically deploys a trained SNN across a commodity CPU and its on-chip dense accelerator, and no prior work shows how the arrangement of spike activity determines which engine is preferable.

We study the EFR32MG24 MCU, which combines a Cortex-M33 CPU with an on-chip Matrix Vector Processor (MVP), to ask whether the MVP can outperform optimized sparse CPU execution of the same SNN. Our automated snnTorch-to-C flow uses spike activity profiles, measured per-layer costs, and memory constraints to select an execution engine for each affine layer. Because accelerator efficiency depends on the arrangement as well as the amount of spike activity, we also introduce function-preserving neuron reordering that places co-active neurons adjacently, improving MVP locality without retraining or changing network function.

Layer-level mapping alone, however, leaves the Cortex-M33 idle during accelerator execution. We therefore pipeline stateless affine operations across timesteps to overlap MVP affine execution at timestep $t$ with Cortex-M33 neuron-state updates from preceding timesteps, and account for it in engine selection, without changing outputs. Together, the deployment flow, activity-aware reordering, and asynchronous scheduling address the gaps left by prior sparse CPU deployments, SNN frameworks, and dense accelerator toolchains.

\begin{figure*}[t]
    \centering
    \includegraphics[width=1\textwidth]{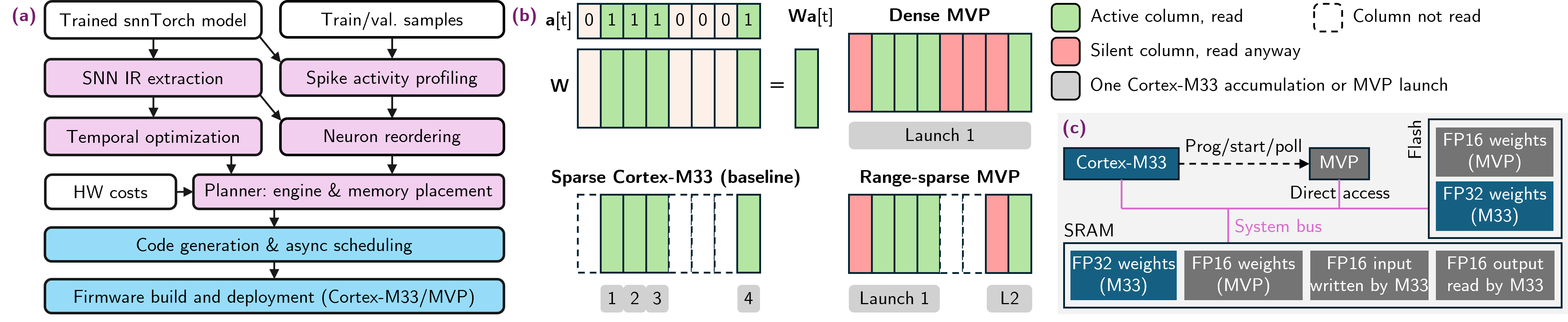}
    \caption{(a) Deployment flow. (b) One affine layer under sparse Cortex-M33, dense MVP, and range-sparse MVP execution. (c) Memory organization: the MVP accesses SRAM and Flash directly; the Cortex-M33 programs, starts, and polls the MVP through memory-mapped registers.}
    \label{fig:flow}
\end{figure*}

This work makes three contributions. \textbf{First}, we develop, to our knowledge, the first automated snnTorch-to-C deployment flow that maps trained SNNs across a commodity MCU's CPU and on-chip dense accelerator. On EFR32MG24 hardware, heterogeneous Cortex-M33--MVP execution achieves 1.27--1.87$\times$ speedup over optimized sparse CPU inference on five of six evaluated network configurations, while the sixth retains CPU-optimal execution. \textbf{Second}, we show that function-preserving hidden-neuron reordering can change the preferred execution engine of a trained network layer without retraining or altering spike activity. \textbf{Third}, we develop dependency-aware asynchronous scheduling that overlaps Cortex-M33 neuron-state updates with MVP affine execution, improving end-to-end latency beyond what blocking heterogeneous mapping alone achieves.

\section{SNN Model and Workloads}

\subsection{Discrete-Time SNN Model}

We consider feed-forward SNNs with leaky integrate-and-fire (LIF) neurons following the discrete-time snnTorch formulation~\cite{eshraghian2023training}. For layer $l$ at timestep $t$, $\mathbf{s}^{(l-1)}[t]$ is the input spike vector, $\mathbf{W}^{(l)}$ and $\mathbf{b}^{(l)}$ are the learned affine parameters, and $\mathbf{u}^{(l)}[t]$ is the membrane state. With decay $\boldsymbol{\beta}^{(l)}$, threshold $\boldsymbol{\theta}^{(l)}$, unit-step function $H(\cdot)$, and element-wise product $\odot$,
\begin{equation}
\begin{aligned}
\mathbf{i}^{(l)}[t]&=
\mathbf{W}^{(l)}\mathbf{s}^{(l-1)}[t]+\mathbf{b}^{(l)},\\
\tilde{\mathbf{u}}^{(l)}[t]&=
\boldsymbol{\beta}^{(l)}\odot\mathbf{u}^{(l)}[t-1]+\mathbf{i}^{(l)}[t],\\
\mathbf{s}^{(l)}[t]&=
H\!\left(\tilde{\mathbf{u}}^{(l)}[t]-\boldsymbol{\theta}^{(l)}\right),\\
\mathbf{u}^{(l)}[t]&=
\tilde{\mathbf{u}}^{(l)}[t]-\boldsymbol{\theta}^{(l)}\odot\mathbf{s}^{(l)}[t].
\end{aligned}
\label{eq:lif}
\end{equation}

The affine computation carries no recurrent state, whereas the LIF update maintains membrane state across timesteps. We therefore map affine computation to the Cortex-M33 or MVP independently per layer, while neuron dynamics remain on the Cortex-M33.

\subsection{Workloads}
\label{sec:workloads}

We evaluate three event-based classification workloads. Spiking Heidelberg Digits (SHD) uses a $700\!-\!64\!-\!20$ network and Spiking Speech Commands (SSC) a $700\!-\!64\!-\!64\!-\!35$ network, both with $T=100$ inference timesteps~\cite{cramer2022heidelberg}. Their sparse 700-channel external inputs test whether naturally occurring input locality can make the first affine layer favorable for MVP execution.

Braille~\cite{mullercleve2022braille} uses 24 event channels and $T=51$ with $24\!-\!192\!-\!192\!-\!27$ and $24\!-\!384\!-\!384\!-\!27$ networks. Here, the main accelerator opportunity comes from hidden-layer LIF spike activity, which lets us study how post-training neuron reordering affects the choice between MVP and Cortex-M33 execution.

To test whether the reordering-induced crossover generalizes beyond Braille, we train SSC networks with hidden widths 128 and 192 using the same dataset, timestep count, and training procedure as the 64-wide model. Unlike SHD, SSC has a hidden-to-hidden Linear1, which grows in both dimensions with width.

For all networks, affine layers are numbered in execution order: Linear0 maps the external input to the first hidden layer, Linear1 maps the first hidden layer to the next layer (hidden or output), and Linear2, where present, is the final readout.

\section{Heterogeneous Sparse Execution}

\subsection{Deployment Flow}

As shown in Fig.~\ref{fig:flow}(a), our snnTorch-to-C deployment flow\footnote{SpikeDuet is available at \url{https://github.com/chaselab-lums/spikeduet}.} first extracts the trained model into an SNN intermediate representation (IR) and profiles per-layer spike activity on two splits: training-split activity drives function-preserving neuron reordering (Section~\ref{sec:reorder}) and hot-column selection (Section~\ref{sec:memory}), while validation-split profiles inform engine selection. The IR also undergoes temporal optimization. The planner combines the validation-split profiles with hardware cost models and the 256~KiB SRAM budget of the EFR32MG24~\cite{silabs_efr32xg24_rm} to fix each affine layer's execution engine and weight residency (Section~\ref{sec:engine}). Finally, the flow generates C code with asynchronous CPU--MVP scheduling (Section~\ref{sec:async}) and builds the firmware.

\subsection{Sparse Affine Execution and Engine Selection}
\label{sec:engine}

The MVP fetches FP16 weights from SRAM or Flash and an FP16 input vector written by the Cortex-M33, either cast from FP32 external inputs or written directly by the LIF update for hidden spikes, and writes its output back without CPU data movement; the Cortex-M33 only programs, starts, and polls each launch (Fig.~\ref{fig:flow}(c)). For an affine layer with input $\mathbf{a}[t]$, weights $\mathbf{W}$, and bias $\mathbf{b}$,
\begin{equation}
\mathbf{W}\mathbf{a}[t]+\mathbf{b}
=
\sum_{j:a_j[t]\neq0}\mathbf{W}_{:,j}a_j[t]+\mathbf{b},
\label{eq:sparse_affine}
\end{equation}
where $\mathbf{W}_{:,j}$ is column $j$ of $\mathbf{W}$. Fig.~\ref{fig:flow}(b) illustrates the three resulting mappings. \emph{Sparse Cortex-M33} execution, our baseline, accumulates only the columns of active inputs. Its cost scales with the number of active inputs and output neurons, and depends on weight residency: SRAM-resident accumulation costs about 2.5 cycles per active input per output neuron, and Flash-resident weights roughly double this.

\emph{Dense MVP} execution processes every column, including silent ones, in a single matrix-vector operation. The MVP favors such regular, contiguous accesses~\cite{silabs_efr32xg24_rm}, but launching it once per active column would incur excessive overhead. Our \emph{range-sparse MVP} kernel sits between these extremes: it groups neighboring FP16 column pairs into contiguous ranges, bridging inactive gaps when cheaper than an additional launch. Its measured per-timestep cost is modeled as
\begin{equation}
C_{\mathrm{MVP,sparse}}[t]
=
C_0[t]+C_LN_L[t]+C_PN_P[t],
\label{eq:sparse_cost}
\end{equation}
where $N_L$ counts accelerator launches (one per range), $N_P$ the column pairs processed, including bridged inactive pairs, and $C_0$ a layer-dependent overhead. For the 64-output input layer, $C_L\!\approx\!454$ cycles per launch and $C_P\!\approx\!88$ cycles per pair, including input casting; other output widths use coefficients calibrated on hardware. For hidden-spike inputs, the FP16 write falls in $C_0$. Per timestep, the kernel uses a single launch over all columns when cheaper than launching each range.

\emph{Heterogeneous execution} assigns each affine layer one of these three mappings. The planner runs offline on validation-split profiles and picks, for each affine layer, the cheapest combination of engine and weight residency within the SRAM budget, accounting for CPU--MVP overlap under asynchronous scheduling. Heterogeneous builds are evaluated with both blocking and asynchronous CPU--MVP scheduling, and share code generation, profiles, and memory and temporal optimizations with the baseline.

\subsection{Function-Preserving Neuron Reordering}
\label{sec:reorder}

The range-sparse cost depends on which active columns are adjacent, and hence on the order of hidden neurons, which is functionally arbitrary: permuting a hidden layer and applying the inverse permutation to the following layer preserves the network function. Unlike prior uses of this symmetry for model merging and pruning~\cite{entezari2022permutation,wan2023upscale}, we exploit it to place co-active neurons adjacently. For permutation matrix $\mathbf{P}$, $\mathbf{W}^{(l)\prime}=\mathbf{P}\mathbf{W}^{(l)}$ and $\mathbf{W}^{(l+1)\prime}=\mathbf{W}^{(l+1)}\mathbf{P}^{T}$; biases, neuron parameters, and runtime state are permuted consistently, so that $\mathbf{s}^{(l)\prime}[t]=\mathbf{P}\mathbf{s}^{(l)}[t]$ and $\mathbf{W}^{(l+1)\prime}\mathbf{s}^{(l)\prime}[t]=\mathbf{W}^{(l+1)}\mathbf{s}^{(l)}[t]$.
Unlike activity tuning~\cite{barchi2024activity}, reordering changes neither which neurons fire nor the sparse Cortex-M33 work, only the adjacency of active columns seen by the MVP. We derive neuron co-activity from the training split, obtain an initial ordering by hierarchical clustering, and refine it with local swaps that minimize the hardware-calibrated cost $C_LN_L+C_PN_P$; $C_0$ is omitted as it is permutation-invariant.

\subsection{Memory and Temporal Optimizations}
\label{sec:memory}

Engine selection is coupled to memory placement because the two execution paths use different weight formats. Sparse Cortex-M33 execution benefits from SRAM-resident weights; for Flash-resident FP32 tables, as in Braille-384, SSC-128, and SSC-192, frequently accessed \emph{hot columns} are identified from training-split activity and copied into SRAM at startup. The MVP instead accesses contiguous FP16 weight ranges directly from SRAM or Flash, without hot-column pinning.

Both Braille models use a non-resetting linear readout whose outputs are averaged over time during training, with the predicted class given by the largest output. Since averaging and summing over time select the same class, and the readout is linear, $\sum_t(\mathbf{W}_{\mathrm{out}}\mathbf{s}[t]+\mathbf{b}_{\mathrm{out}})=\mathbf{W}_{\mathrm{out}}\sum_t\mathbf{s}[t]+T\mathbf{b}_{\mathrm{out}}$, so the firmware evaluates the readout once on accumulated spike counts rather than at every timestep.

\subsection{Asynchronous CPU--MVP Execution}
\label{sec:async}

In blocking execution, the Cortex-M33 waits for each MVP launch to finish. We instead overlap MVP affine execution at timestep $t$ with Cortex-M33 LIF updates and readout from preceding timesteps. Nonblocking launches are chained by cooperative polling, reducing MVP idle time while the CPU does useful work. Separate intermediate buffers and a final synchronization preserve data dependencies and outputs, and the per-timestep input interface is unchanged.

\section{Experimental Evaluation}

\subsection{Experimental Setup and Validation}

All measured latencies use the EFR32MG24 at 39~MHz and the Cortex-M33 Data Watchpoint and Trace cycle counter; energy is computed from the measured supply current at 3.3~V and the inference latency. End-to-end comparisons use fixed test sets of 40 SHD inputs, 70 inputs per SSC network, and 108 inputs per Braille network. Layer-level Linear1 measurements use validation inputs, which also feed engine-selection profiling: 108 (four per class) for Braille-192 and Braille-384, and ten per model for SSC-128 and SSC-192.

Firmware outputs were validated against the same generated code compiled and run on a host PC. Sparse kernels produced zero bitwise mismatches against dense affine recomputation over 4,000 SHD and 7,000 SSC-64 timesteps. We also emulate the deployed numerical formats in software, FP32 on the Cortex-M33 and FP16 on the MVP. Relative to the FP32 reference, accuracy on the full test splits changes by $-0.49$, $-0.05$, $+0.15$, $+0.05$, and $-0.12$ percentage points on SHD, SSC-64, SSC-128, SSC-192, and Braille-384, respectively.

\subsection{Measured Heterogeneous Benefit}
\label{sec:het-benefit}

\begin{table}[t]
\centering
\caption{Measured end-to-end inference latency (ms) and energy}
\label{tab:async}
\scriptsize
\setlength{\tabcolsep}{2pt}
\renewcommand{\arraystretch}{1.1}
\begin{adjustbox}{max width=\columnwidth}
\begin{tabular}{@{}lrrrrrcr@{}}
\toprule
\textbf{Workload} & \textbf{M33} & \makecell{\textbf{Dense}\\\textbf{MVP}} & \textbf{Block.} & \textbf{Async} & \textbf{Gain} & \makecell{\textbf{MVP layer}\\\textbf{(async)}} & \makecell{\textbf{Energy (mJ)}\\\textbf{M33/Async}} \\
\midrule
SHD         & 48.75  & 83.93  & 41.61  & 37.98  & 1.28$\times$ & L0 & 0.37/0.31 ($-$16\%) \\
SSC-64      & 59.28  & 100.56 & 55.37  & 46.67  & 1.27$\times$ & L0 & 0.44/0.39 ($-$11\%) \\
SSC-128     & 131.85 & 190.79 & 94.09$^\ddagger$  & 84.11  & 1.57$\times$ & L0 & 0.97/0.74 ($-$23\%) \\
SSC-192     & 209.40 & 390.47 & 139.50$^\ddagger$ & 112.07 & 1.87$\times$ & L0 & 1.53/1.01 ($-$34\%) \\
Braille-192 & 31.26  & 55.64  & 31.26  & --     & 1.00$\times$ & -- & 0.24 (0\%) \\
Braille-384 & 89.93  & 208.53 & 82.78$^\dagger$  & 66.14  & 1.36$\times$ & L1$^\dagger$ & 0.64/0.56 ($-$13\%) \\
\bottomrule
\end{tabular}
\end{adjustbox}
\parbox{\columnwidth}{\justifying\scriptsize
Blocking and asynchronous columns report the best evaluated configurations; gain and per-inference energy are relative to sparse M33. L0 and L1 denote Linear0 and Linear1; Braille-192 uses no MVP layer. $^\ddagger$Blocking maps reordered L1 to the MVP; asynchronous execution retains the trained order with only L0 on the MVP. $^\dagger$Uses reordered L1.}
\end{table}

Table~\ref{tab:async} summarizes measured end-to-end latency across six networks. Dense MVP execution is 1.4--2.3$\times$ slower than sparse Cortex-M33 execution throughout, whereas asynchronous heterogeneous execution achieves 1.27--1.87$\times$ speedup on all networks except Braille-192, which retains sparse Cortex-M33 execution (Section~\ref{sec:crossover}). For Braille-384, with engine mapping and weight residency fixed (Linear1 on the range-sparse MVP with Flash-resident FP16 weights), reordering reduces blocking latency from 97.01 to 82.78~ms (14.7\%) and asynchronous latency from 85.19 to 66.14~ms (22.4\%), so its benefit persists under CPU--MVP concurrency; combined, they yield the 1.36$\times$ speedup over the hot-column sparse M33 baseline. Measured energy savings range from 11\% to 34\% across the five accelerated workloads, with the largest reduction on SSC-192.

\begin{figure}[t]
\centering
\includegraphics[width=1.0\columnwidth]{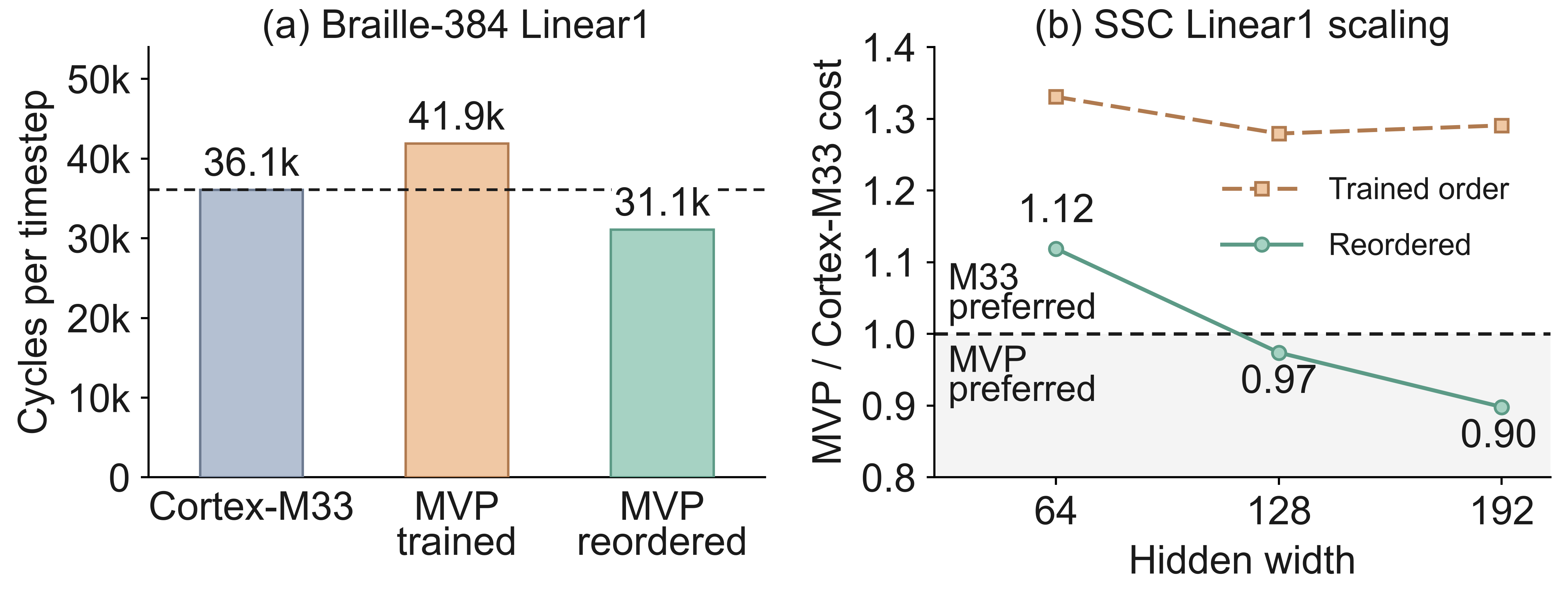}
\caption{Activity-locality effects on execution-engine selection. (a) Measured Braille-384 Linear1 execution cycles before and after neuron reordering, demonstrating an M33-to-MVP crossover. (b) Hardware-calibrated SSC Linear1 MVP-to-M33 cost ratios after reordering across hidden-layer widths; ratios below unity favor MVP execution.}
\label{fig:engine_crossover}
\end{figure}

\subsection{Locality-Induced Engine Crossover}
\label{sec:crossover}

Figure~\ref{fig:engine_crossover}(a) shows the measured Braille-384 Linear1 crossover on validation inputs not used to choose the neuron order (fitted on the training split). Reordering groups co-active neurons into fewer, longer ranges, cutting the average number of MVP launches per timestep from 10.1 to 5.8, while spike activity and Cortex-M33 work stay unchanged.

Sparse Cortex-M33 execution with hot columns averages 36,101 cycles per timestep, compared with 41,916 cycles for MVP execution in the original neuron order (1.16$\times$ the M33 cost). After reordering, MVP execution falls to 31,123 cycles (0.86$\times$). Across all evaluated validation inputs, the MVP is slower than the M33 before reordering and faster afterward. Changing activity locality alone thus reverses the preferred execution engine for the same trained network.

Braille-192 provides a measured counterexample. Reordering reduces its average Linear1 range count per timestep from 3.49 to 2.19, yet MVP execution requires 7,383 cycles versus 5,604 on the sparse M33 (1.32$\times$). Improved locality alone is therefore insufficient to make this smaller layer accelerator-favorable.

\subsection{SSC Width Scaling: Predictions and Hardware Validation}

Across SSC hidden widths of 64, 128, and 192 (Section~\ref{sec:workloads}), mean first-hidden-layer spike density decreases from 12.2\% to 8.8\% and 7.1\%, respectively. Because learned activity and accuracy also vary with width, this analysis is treated as corroborating evidence rather than an isolated width-scaling experiment.

\textbf{Layer-level crossover.} In the trained neuron order, hardware-calibrated models predict that range-sparse MVP execution of Linear1 costs 28--33\% more than sparse Cortex-M33 execution across the three widths. After reordering, the predicted MVP-to-M33 cost ratios are 1.12, 0.97, and 0.90 for widths 64, 128, and 192, respectively (Fig.~\ref{fig:engine_crossover}(b)). Hardware measurements on validation inputs, with SRAM-resident weights on both engines and the FP16 spike write included, give ratios of 1.02 at width 128 (5,507 vs.\ 5,388 cycles per timestep) and 0.83 at width 192 (8,088 vs.\ 9,698 cycles). Individual predicted engine costs are within 6\% of measurements. Thus, Linear1 approaches break-even at width 128 and becomes accelerator-favorable at width 192.

\textbf{End-to-end effects.} Across three blocking configurations, the models predict end-to-end latency within 4.1\% for SSC-128 and 1.7\% for SSC-192. Reordering lowers blocking latency on both, but for different reasons. On SSC-192 (143.12 to 139.50~ms, 2.5\%), the gain comes from the Linear1 crossover. On SSC-128 (101.97 to 94.09~ms, 7.7\%), it comes from memory placement: in the trained order, the M33's FP32 Linear1 and Linear2 tables (82~KiB) leave no room for the MVP's FP16 Linear0 weights (175~KiB), whereas moving Linear1 to the MVP removes its FP32 table so Linear0 fits in SRAM, saving 8.0~ms while Linear1 becomes 0.3~ms slower. Under asynchronous scheduling, the best configurations keep the trained order with only Linear0 on the MVP (Table~\ref{tab:async}), since leaving Linear1 on the M33 lets it overlap with Linear0's MVP execution, even at width 192, where Linear1 is cheaper on the MVP alone. Engine selection thus depends on activity locality, memory residency, and concurrency; Braille-384 shows their combined benefit (Section~\ref{sec:crossover}).

\section{Conclusion}

This work shows that commodity matrix-vector accelerators can complement optimized sparse CPU execution of SNNs. 
On the EFR32MG24, heterogeneous Cortex-M33--MVP execution achieves 1.27--1.87$\times$ speedup and 11--34\% energy savings over sparse Cortex-M33 inference on five of six evaluated networks, while the smallest remains CPU-optimal. Accelerator suitability depends not only on spike sparsity but also on activity locality: function-preserving neuron reordering, without retraining or altering spike activity, moves the dominant Braille-384 hidden layer from Cortex-M33 to MVP execution, and asynchronous scheduling further reduces latency by overlapping CPU and MVP work. As SSC-128 and SSC-192 show, the best engine choice also depends on whether CPU and MVP execution can overlap. Future work will extend the flow to other accelerator-equipped MCUs.

\section*{Acknowledgment}
The authors used Claude (Anthropic) for grammar and text refinement; all technical content, figures, results, and claims were produced and verified by the authors.

\bibliographystyle{IEEEtran}
\bibliography{references}

\end{document}